\documentclass[a4paper]{jpconf}
\usepackage{graphicx}

\def\ltap{\ \raise.3ex\hbox{$<$\kern-.75em\lower1ex\hbox{$\sim$}}\ }
\def\gtap{\ \raise.3ex\hbox{$>$\kern-.75em\lower1ex\hbox{$\sim$}}\ }

\begin{document}
\title{Charged-Current and Neutral-Current Coherent Pion Productions\ \
\  ---\ \    Theoretical Status}

\author{Satoshi X. Nakamura}

\address{Excited Baryon Analysis Center, Jefferson Lab, Newport News,
VA, 23606, USA}

\ead{satoshi@jlab.org}

\begin{abstract}
Theoretical status of CC and NC coherent pion productions is reviewed.
\end{abstract}

\section{Introduction}

Recent neutrino oscillation experiments have driven quite a few
nuclear theorists to study neutrino-nucleus interaction.
Theoretical study of coherent pion productions (coh$\pi$) has been also on the
trend.
The charged-current (CC) coh$\pi^+$ is 
$\nu_l + A_{g.s.}\to l^- + A_{g.s.} + \pi^+$, while
the neutral-current (NC) coh$\pi^0$ is 
$\nu_l + A_{g.s.}\to \nu_l + A_{g.s.} + \pi^0$,
where $\nu_l$ is the neutrino of $l=e,\mu,\tau$, and $A_{g.s.}$ is a
nucleus in its ground state.
The process takes place under small momentum transfer from the lepton.
The NC coh$\pi^0$ needs to be
understood for the $\nu_e$ appearance event, because when one of two
photons from $\pi^0\to2\gamma$ is not detected, the NC coh$\pi^0$ can
fake the $\nu_e$ appearance event.
Theoretically, there are two main approaches to coh$\pi$.
One of them is based on the partially conserved axial
current (PCAC)~\cite{RS,RS2,hernandez,Kartavtsev,BS,paschos}, 
and
the other is based on a microscopic dynamical 
model~\cite{kelkar,singh,AR,AR2,amaro,martini,sxn,hernandez2}.
In the following, I discuss both of the approaches, and then comparison among
them and with recent experimental data.

\section{PCAC-based models}

The prominent PCAC-based model was proposed by Rein and
Sehgal~\cite{RS} (RS model). 
The model is based on the fact that a coh$\pi$ amplitude is
proportional to the divergence of the axial-current
in the limit of $q^2\to 0$ where $q$ is the momentum transfer from the
lepton. 
The PCAC dictates that the divergence of the axial-current is
proportional to the pion field. 
Thus, the coh$\pi$ amplitude and the elastic $\pi$-nucleus scattering
amplitude is related. 
Continuation of the cross section to $q^2\ne 0$ is parametrized solely by a
dipole-type function.
The elastic pion-nucleus cross section is simply given by the product of the
pion-nucleon elastic cross section, the square of the nuclear form
factor, and the pion absorption factor.
The RS model works well for high
energy neutrino ($E_\nu\gtap$ 2~GeV), from medium to heavy nuclei~\cite{charm}.
The RS model has been used in many analyses of neutrino 
experiments because of its success and simplicity.

However, recent neutrino experiments use neutrino beam of
$E_\nu\ltap$ 2~GeV and relatively light nuclei ({\it e.g.}, $^{12}$C) as a target,
for which the validity of the RS model is questionable.
Indeed, recent experiments~\cite{hasegawa,hiraide} found that CC
coh$\pi^+$ cross section of the RS model is significantly larger than
their data.
Rein and Sehgal corrected the RS model by taking account of the finite
muon mass whose effect can be significant for low $E_\nu$~\cite{RS2}.
They found 25\% reduction of the CC coh$\pi^+$ cross sections for 
$E_\nu = 1.3$~GeV.
Further improvements were attempted by Hern\'andez {\it et al.}~\cite{hernandez}
They pointed out that the elastic pion-nucleus scattering model in the
RS model is rather off data.
They improved the model by considering previously ignored $t$-dependence
of the pion-nucleon cross sections, 
and by using a more realistic pion absorption factor.
It was found that, although
the pion-nucleus elastic scattering is significantly improved, 
the coh$\pi$ cross sections 
from the improved RS model are still rather
different from those of microscopic calculations.

Meanwhile, another avenue for the PCAC-based model was explored~\cite{Kartavtsev,BS,paschos}.
There, the PCAC relation was used to relate the axial current-nucleus
interaction to the pion-nucleus interaction (nuclear PCAC), 
rather than the axial current-nucleon
interaction to the pion-nucleon interaction (nucleon PCAC); the latter was employed in
the RS model.
Thus, they can use the pion-nucleus cross section data directly to 
calculate the coh$\pi$ cross sections.
They found that this change yields a drastic reduction ($\sim 1/2$) of
the coh$\pi$ cross sections,  now the size is comparable to those of
the microscopic calculations.
Strictly speaking, however, 
the nuclear PCAC relates a coh$\pi$ amplitude to the half
off-shell pion-nucleus scattering amplitude~\cite{bell}.
Therefore, the use of pion-nucleus cross section may call for some
justification.

\section{Dynamical microscopic models}

A description of coh$\pi$ with a dynamical microscopic model consists of
three basic ingredients: elementary amplitudes (neutrino-induced pion
production off a single nucleon); nuclear medium effect on
the $\Delta$-propagation; pion-nucleus final state interaction.
In the following, I will discuss each of them used in the previous
works.

\vspace{2mm}
\noindent{\it Elementary amplitude} : 
The most important mechanism for the pion production is the
$\Delta$-excitation.
Thus Refs.~\cite{kelkar,singh,AR} used only the $s$-channel $\Delta$
diagram for the elementary amplitude. 
Among the $AN\Delta$ ($A$:axial-current; $N$:nucleon) form factors, the so-called 
$C_5^{A} (q^2)$ gives a dominant contribution so that the coh$\pi$ cross
section is roughly $\sigma_{\rm coh\pi} \propto |C_5^A|^2$.
In Refs.~\cite{kelkar,singh,AR},  the Goldberger-Treiman (GT) value for
$C_5^A(0)$ was taken:
$C_5^A(0) = g_{\pi N\Delta} f_\pi/\sqrt{6} m_N$ = 1.2.
Martini {\it et al.}~\cite{martini} used $s$- and $u$-channel
nucleon and $\Delta$ mechanisms as the elementary amplitudes;
the value of $C_5^A(0)$ is the GT value.
Meanwhile, elementary amplitudes used in Refs.~\cite{AR2,amaro,hernandez2}
consists of $s$- and $u$-channel nucleon and $\Delta$ mechanisms,
$t$-channel $\pi$-exchange mechanism, a contact and a pion pole terms~\cite{hnv}.
Because of the background terms, the value of $C_5^A(0)$ has to be
smaller than the GT value by 20\% in order to fit neutrino data.
So far, all elementary amplitudes are based on tree-level calculations.
On the other hand, Sato {\it et al.} proposed a model (SL model~\cite{sl}) in
which tree-level weak
pion production mechanisms, which are similar to those of
Ref.~\cite{hnv}, are dressed by multiple $\pi N$ rescattering to yield
a unitary elementary amplitude.
The SL amplitude contains significant pion cloud effect, which is unique
in this model.
Bare $AN\Delta$ couplings were fixed using the constituent quark model.
Our group used the elementary amplitudes from the SL model~\cite{sxn}.
All elementary amplitudes used in various groups fairly reproduce
available data.
Unfortunately, available data are rather limited, and not enough to test
each elementary amplitude model.
More good quality data would help.

\vspace{2mm}
\noindent{\it Medium effect on $\Delta$} : 
In the vacuum, the $\Delta$ propagates by repeating the decay into 
$\pi N$ state and formation of $\Delta$ (self energy).
In a nuclear medium, this process becomes more complicated.
Some $\pi N$ states are forbidden by the Pauli blocking so that the $\Delta$
decay is suppressed. 
A pion emitted from the $\Delta$ can be absorbed by another nucleon, forming
multiparticle-multihole configurations or 
RPA(random phase approximation)-type processes.
The $\Delta$ mass and width can shift due to these medium effects.
Oset {\it et al.} calculated the medium effects on the $\Delta$ propagation
using a many-body approach~\cite{oset}, and parametrized the shift of the
$\Delta$ mass and width as a function of the density.
Their result has been employed by many people for calculating the
coh$\pi$ cross sections~\cite{kelkar,singh,AR,AR2,amaro,martini,hernandez2}.
However, non-locality due to the recoil of the $\Delta$ was not
explicitly taken into account in these calculations of coh$\pi$. 
Leitner {\it et al.}~\cite{leitner} pointed out this, and showed that the
non-local effect can reduce the total coh$\pi$ cross sections by half.
Hern\'andez {\it et al.} argued that the non-local effect is effectively (partially)
taken into account in the course of fitting the medium shift of the
$\Delta$ to observables~\cite{hernandez2}.
Our group~\cite{sxn} considered the non-locality of the $\Delta$-propagation
as well as several medium effects explicitly. 
Multiparticle-multihole effects are simulated by a 
phenomenological spreading potential, parameters of which were fitted to
pion-nucleus elastic and total (elastic+inelastic) cross sections.
It is important to fit the total cross sections because the spreading
potential describes pion absorptions.

\vspace{2mm}
\noindent{\it Final state interaction (FSI)} : 
In Refs.~\cite{kelkar,AR,AR2,amaro,hernandez2}, pion-nucleus FSI is
considered in distorted pion wave function obtained by solving the
Klein-Gordon equation with an optical potential.
The optical potential is based on the $\Delta$-hole model and
the Lorentz-Lorentz correction, and reproduces spectra of pionic atoms
and low-energy pion-nucleus elastic cross sections~\cite{pot}.
In Martini {\it et al.}'s approach~\cite{martini}, the elastic pion-nucleus
cross section is obtained by solving a RPA equation with a bare
polarization propagator as the driving term.
The bare polarization propagator contains various nuclear excitation
such as multiparticle-multihole configurations.
The model well describes the elastic total cross sections in the $\Delta$ region.
Our group constructed a pion-nucleus optical potential from $\pi N$
amplitudes of the SL model, combined with the medium effects on the
$\Delta$. A pion-nucleus scattering amplitude is obtained by
solving the Lippmann-Schwinger equation with the optical potential.
The model well describes the elastic differential cross sections and the
total cross sections in the $\Delta$ region.

\begin{figure}[t]
\begin{minipage}[c]{77mm}
\begin{center}
      \includegraphics[width=72mm]{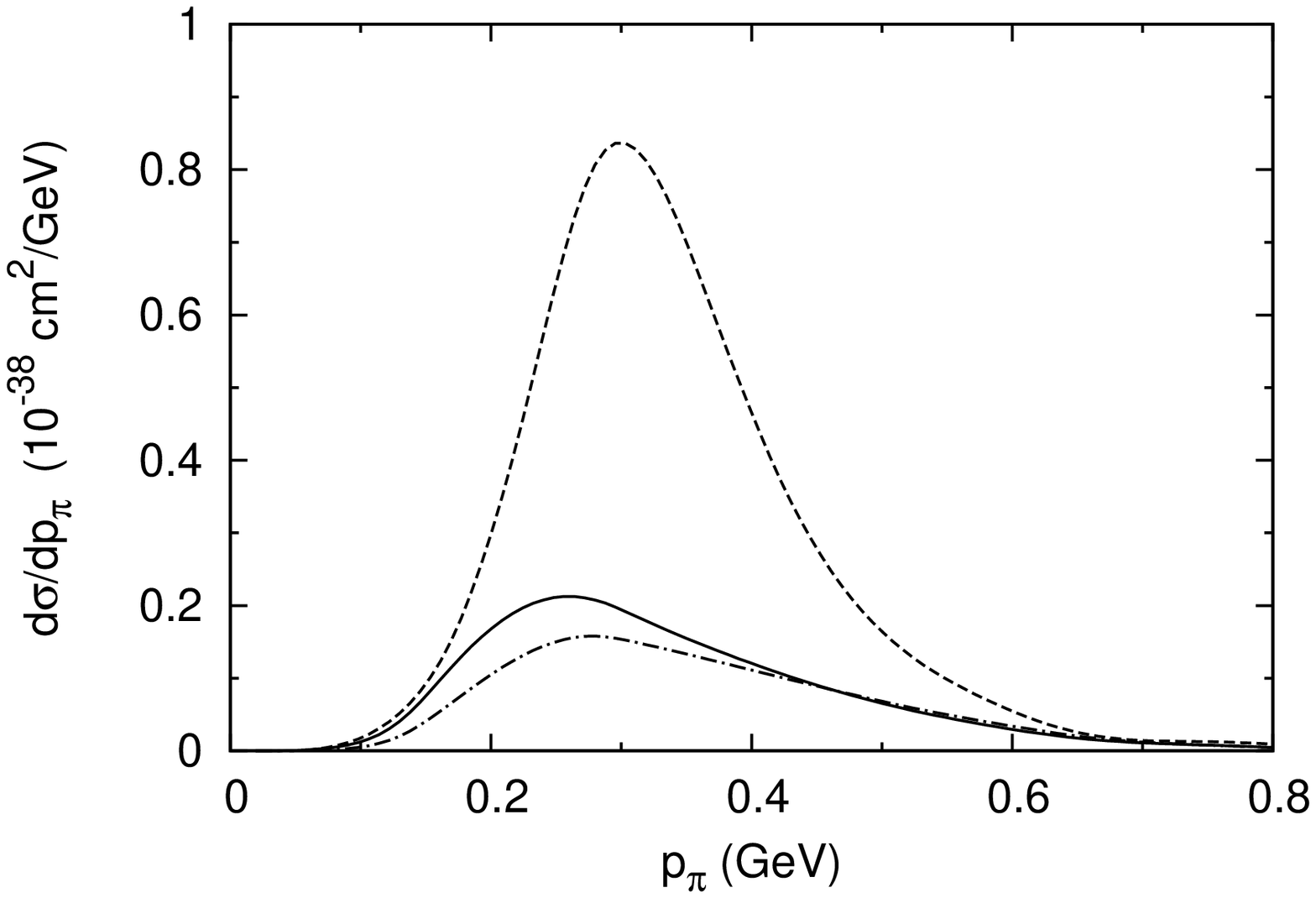} 
\caption{\label{fig:coh_cc} Pion momentum distribution in CC coh$\pi$ at
 $E_\nu=1$~GeV from Ref.~\cite{sxn}. See the text for more.
}
\end{center}
\end{minipage}
\hspace{2mm}
\begin{minipage}[c]{77mm}
      \includegraphics[width=7cm]{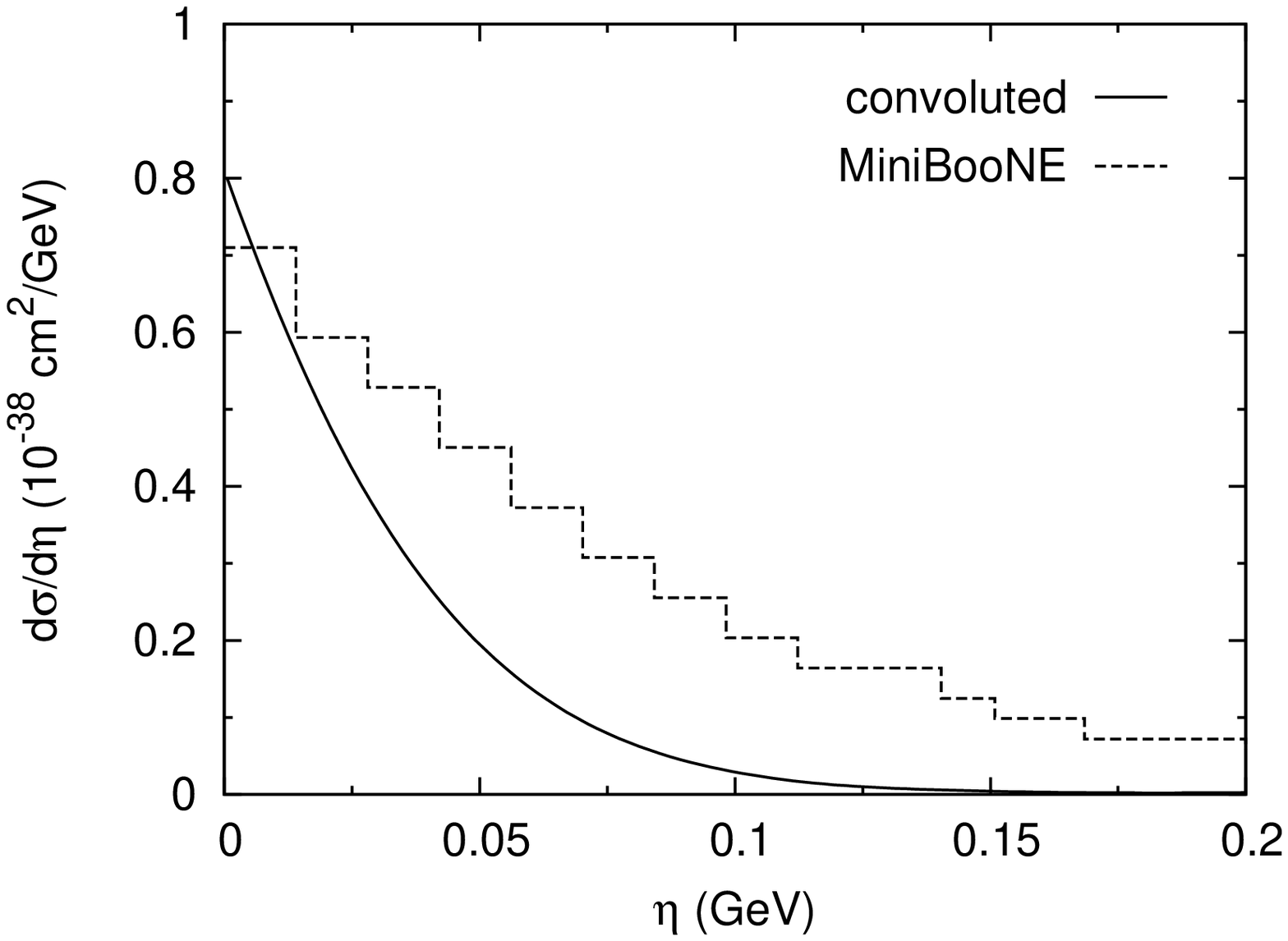}
\caption{\label{fig:eta}
$\eta$-distribution for NC coh$\pi^0$ on $^{12}$C.
Solid: model calculation~\cite{sxn} convoluted with MiniBooNE flux;
Dash: MC output.
}
\end{minipage}
\end{figure}

\vspace{2mm}
\noindent{\it Coherent pion production} :
Combining the above-discussed three ingredients, one can calculate coh$\pi$ cross
sections.
First let us discuss a photo-coh$\pi^0$.
This is a great testing ground for microscopic models of coh$\pi$
because data of good quality are available.
Also, some dynamical models like ours~\cite{sxn} can predict coh$\pi$ cross
sections without any adjustable parameters; all parameters
have been fitted to pion-nucleus scattering data. 
Thus we have tested our prediction for photo-coh$\pi$ with data, and
found a good agreement~\cite{sxn}.
It is noteworthy that the medium effects on $\Delta$ and FSI reduce the
cross sections by about half to reach the good agreement.
This success serves as a good basis to apply our model to the neutrino-coh$\pi$.
The pion momentum distribution in CC coh$\pi^+$ at $E_\nu=1$~GeV is
shown in Fig.~\ref{fig:coh_cc}.
Significant medium effects are observed here.
The dashed curve contain neither the $\Delta$ spreading potential nor
the FSI. Including these nuclear effects lowers the dashed curve into
the solid curve.
The dash-dotted curve contains only resonant amplitude, showing
significant contribution from non-resonant unitary amplitude.
This is in contrast with calculations with tree-level elementary
amplitude~\cite{hnv}, because the non-resonant amplitude hardly contributes.

\section{Comparison among models and data}

At NUINT09, CC and NC coh$\pi$ cross sections from many theoretical models and
Monte Carlo (MC) codes were compared~\cite{boyd}. 
Although details are different, recent theoretical 
models~\cite{Kartavtsev,BS,paschos,AR,AR2,amaro,martini,sxn,hernandez2}
give fairly similar CC and NC coh$\pi$ cross sections. 
However, the MC code outputs, which are based on the RS model with
different modifications, are very different from those of the recent theoretical models.

There have been large discrepancies between experimental data and
theoretical calculations on the coh$\pi$ cross sections.
The latest data from SciBooNE~\cite{kurimoto} gave the ratio between CC coh$\pi^+$ and 
NC coh$\pi^0$: $\sigma_{CC}/\sigma_{NC} = 0.14^{+0.30}_{-0.28}$.
On the other hand, all recent theoretical calculations gave 
$\sigma_{CC}/\sigma_{NC} = 1.5\sim2$, which is expected from the isospin
factor.

One may suspect that this unsatisfactory situation is due to 
using the RS model in the analyses of the NC data.
In fact, several authors~\cite{amaro,sxn,hernandez2} have shown that $\eta$-distribution
$(\eta\equiv E_\pi (1-\cos\theta_\pi))$
of the RS model rather deviates 
from those of their microscopic models (Fig.~\ref{fig:eta}).
The $\eta$-distribution has been used in the analyses of the NC
data in order to decompose $\pi^0$ events into each production
mechanism.
The use of the RS model in the analyses could have overestimated
the NC coh$\pi^0$ cross sections.
It is very interesting to re-analyze the data with a more realistic coh$\pi$ model.

\ack
I thank the NUFACT11 organizers for the invitation
and the financial support for my participation in the workshop.
This work is supported by the U.S. Department of Energy, Office of Nuclear Physics Division, under Contract No. DE-AC05-06OR23177 under which Jefferson Science Associates operates Jefferson Lab.

\end{document}